\documentclass[prl,twocolumn,showpacs,superscriptaddress,10pt]{revtex4-1}
\usepackage{hyperref}
\usepackage{graphicx}
\usepackage{epsfig}
\usepackage{subfigure}
\usepackage{setspace}
\usepackage{amsmath}

\begin{document}

\title{Fine tuning of micropillar cavity modes through repetitive oxidations}

\author{Morten P. Bakker}
\affiliation{Huygens Laboratory, Leiden University, P.O. Box 9504, 2300 RA Leiden, The Netherlands}
\author{Donald J. Suntrup III}
\affiliation{University of California Santa Barbara, Santa Barbara, California 93106, USA}
\author{Henk Snijders}
\affiliation{Huygens Laboratory, Leiden University, P.O. Box 9504, 2300 RA Leiden, The Netherlands}
\author{Tuan-Ahn Truong}
\affiliation{University of California Santa Barbara, Santa Barbara, California 93106, USA}
\author{Pierre M. Petroff}
\affiliation{University of California Santa Barbara, Santa Barbara, California 93106, USA}
\author{Dirk Bouwmeester}
\affiliation{Huygens Laboratory, Leiden University, P.O. Box 9504, 2300 RA Leiden, The Netherlands}
\affiliation{University of California Santa Barbara, Santa Barbara, California 93106, USA}
\author{Martin P. van Exter}
\affiliation{Huygens Laboratory, Leiden University, P.O. Box 9504, 2300 RA Leiden, The Netherlands}

\date{\today}

\begin{abstract}

Repetitive wet thermal oxidations of a tapered oxide aperture in a micropillar structure are demonstrated.
After each oxidation step the confined optical modes are analyzed at room temperature.
Three regimes are identified.
First, the optical confinement increases when the aperture oxidizes towards the center.
Then, the cavity modes shift by more than 30 nm, when the taper starts to oxidize through the center, leading to a decrease in the optical path length.
Finally, the resonance frequency levels off, when the aperture is oxidized all the way through the micropillar, but confined optical modes with a high quality factor remain.
This repetitive oxidation technique therefore enables precise control of the optical cavity volume or wavelength.

\end{abstract}


\maketitle

Semiconductor optical microcavities have found many applications \cite{Vahala2003}.
To obtain optical confinement one commonly used technique is wet thermal oxidation, which makes use of the large difference in effective refractive index between oxidized and unoxidized regions \cite{Baca2005}\cite{Stoltz2005}.
By embedding self-assembled quantum dots (QDs) in etched micropillar cavities combined with an oxide aperture, high-frequency single-photon sources have been realized \cite{Strauf2007}.
Moreover, by combining a large Purcell factor with a high mode-matching efficiency to external fields, such microcavities enabled the study of  cavity quantum electrodynamics in solid state systems \cite{Rakher2009}, which can lead to hybrid quantum information devices \cite{Bonato2010}.

It is a challenge however to obtain reproducible oxidation results, since the wet thermal oxidation process depends strongly on difficult-to-reproduce parameters, such as the roughness of etched sidewalls or the Al composition to within $\sim 1 \%$ accuracy.
One approach is to monitor the formation of oxide layer(s) in real time. This has been demonstrated using several techniques \cite{Sakamoto2002}\cite{Feld1998}\cite{Almuneau2008}\cite{Bakker2013}.
However controlling the size of the oxide aperture, and thereby the exact optical confinement with submicrometer precision, can be challenging by imaging the aperture alone.
An alternative method would then be, once the desired amount of confinement has nearly been reached, to repeat the oxidation in small steps and characterize the device in between until desired optical properties have been achieved.

In this letter we report such a repetitive oxidation technique.
We perform in total 56 oxidation steps of varying duration on the same micropillar and characterize the device after each step at room temperature.
We analyze the optical modes with a resonant reflectance technique to determine the wavelengths of the modes and the modesplitting between fundamental and first-order modes.
Also by scanning a focused 1064 nm laser and monitoring the reflectance across the sample, we obtain an indication of the advancement of the oxide front inside the micropillar.
We demonstrate that the exact optical confinement can be accurately controlled.
Also, by oxidizing the aperture throughout the micropillar, we find that the overall wavelengths of the modes can be tuned by up to 30 nm.

\begin{figure}[b]
\centering
\centerline{\includegraphics[angle=0]{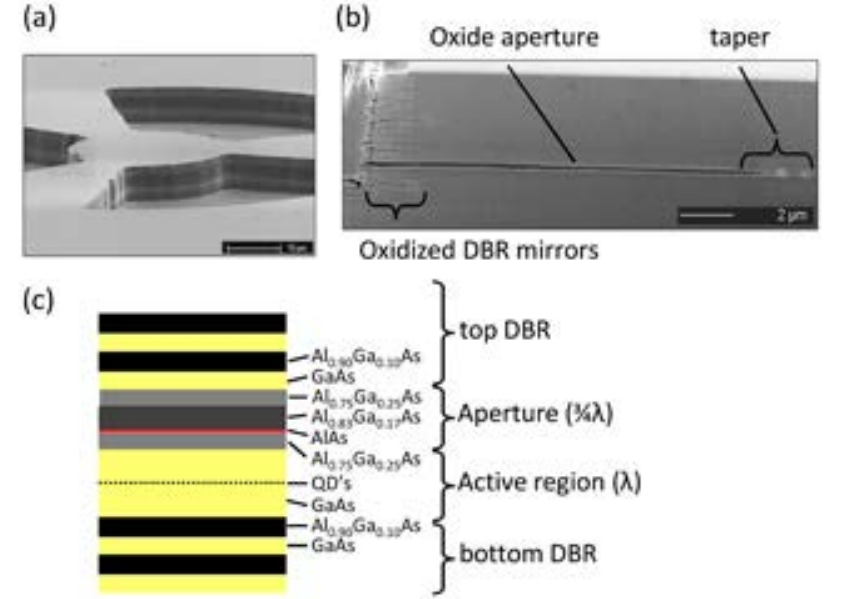}}
\caption{(a) SEM image of the three etched trenches that form a micropillar mesa connected to the bulk material.
(b)SEM image of the FIB cross section from the sidewall of an etched trench to the bulk material. Al$_x$O$_y$ is darker than the Al$_x$Ga$_{1-x}$As layers. [Courtesy to Hozanna Miro (Kavli NanoLab Delft)]
(c) The different Al$_x$Ga$_{1-x}$As layers that make up the aperture and the active region in the center of the micropillar structure. Of the top and bottom DBR mirrors, typically consisting of 30 layer pairs, only a couple $\lambda/4$ layers are shown.
InAs self-assembled quantum dots are located in the center of the GaAs active region in an anti-node of the intracavity intensity, while the AlAs layer in the aperture is located at a node .
}
\label{Fig1}
\end{figure}

The micropillars are constructed as follows \cite{Strauf2007}: first  through molecular beam epitaxy on a GaAs [100] substrate two DBR mirrors are grown, comprising alternating $\lambda/4$-thick layers of GaAs and Al$_{0.9}$Ga$_{0.1}$As. In between the mirrors are a $3/4\lambda$-thick aperture layer, consisting of a thin 10 nm AlAs layer in between Al$_{0.75}$Ga$_{0.25}$As and Al$_{0.83}$Ga$_{0.17}$As, and a $\lambda$-thick active layer containing InGaAs/GaAs self-assembled quantum dots inside a GaAs PIN diode structure.
Then, using reactive ion etching, trenches are etched down to the bottom DBR leaving a 30 $\mu$m circular micropillar connected by three narrow unetched 'spokes' to the bulk material (see Fig. \ref{Fig1}).
Global electrical contacts to the doped layers enable control of the electrical field inside every micropillar in an array of 42 micropillars.

The conversion of Al$_x$Ga$_{1-x}$As to Al$_x$O$_y$ takes place by applying water vapor to the sample heated to 400$^\circ$C.
First we clamp the sample to a holder equipped with a heater, and place it inside a small oxidation furnace equipped with a viewport.
Water vapor is applied by flowing nitrogen through a water-filled bubbler.
While heating up or cooling down the sample in between the oxidations, we purge the furnace with nitrogen in order to prevent dry oxidation through exposure to oxygen.
We speculate that, by preventing exposure to oxygen while the sample is heated, sealing is prevented and repetitive oxidations are thereby made possible.

\begin{figure}
\centering
\centerline{\includegraphics[angle=0]{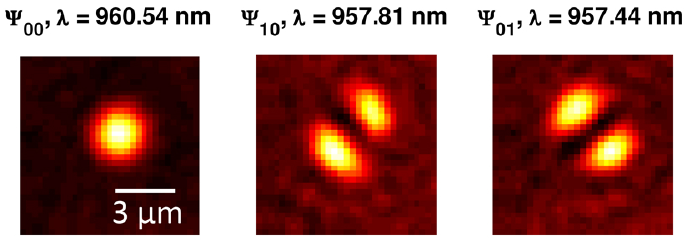}}
\caption{Example of position dependent reflectance measurements, obtained after the 18$^{th}$ oxidation and a total oxidation time of 187 minutes. When a cavity resonance is hit the reflectance decreases (light color). The shape of the modes indicate clearly Hermite-Gaussian modes. The titles denote the labels of the modes and the wavelengths selected from the reflectance spectrometer spectra.}
\label{Fig2}
\end{figure}

At room temperature we characterize the optical modes by scanning a focused superluminescence diode (Superlum, 920-980 nm) over the micropillar center and record the reflectance spectra with a spectrometer \cite{Ctistis2010}.
An example of the spatial reflectance at the different wavelengths corresponding to the fundamental $\Psi_{00}$ and first order $\Psi_{10}/\Psi_{01}$ Hermite Gaussian modes is shown in Fig. \ref{Fig2}.

\begin{figure}
\centering
\centerline{\includegraphics[angle=0]{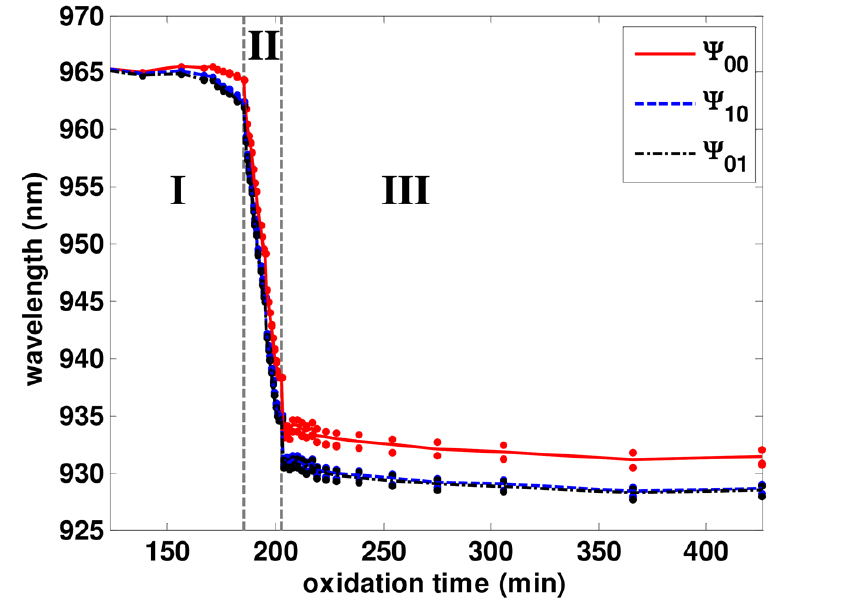}}
\caption{The wavelength of the three lowest order Hermite Gaussian modes, determined at room temperature in between every oxidation step, as function of the total preceding oxidation time. The wavelengths for different polarizations are shown by dots, the average is indicated with a line.}
\label{Fig3}
\end{figure}

In Fig. \ref{Fig3} we present the wavelengths of the three modes as a function of the total time that oxidation has taken place.
A polarization splitting is visible for the different modes due to birefringence.

\begin{figure}
\centering
\centerline{\includegraphics[angle=0]{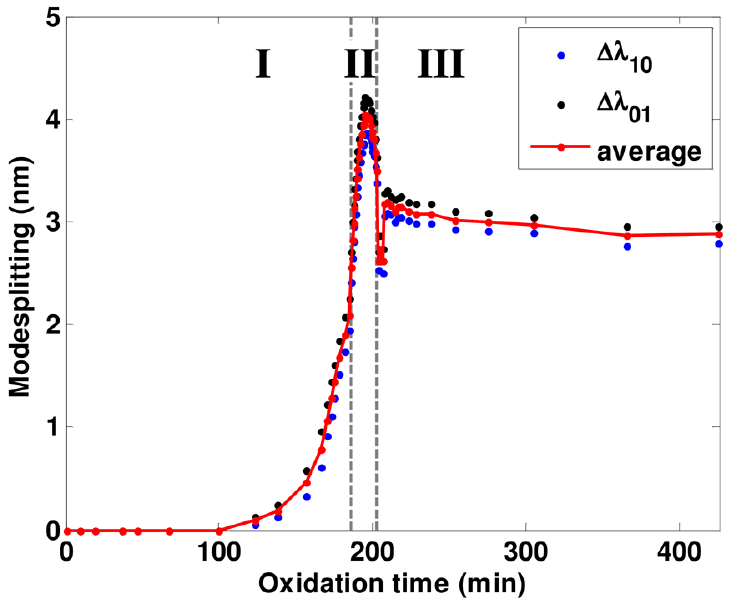}}
\caption{The wavelength splitting between the fundamental and first order Hermite Gaussian modes.}
\label{Fig4}
\end{figure}

From the data we calculate the average mode splittings in Fig. \ref{Fig4}.
A difference in the mode splittings $\Delta\lambda_{10}$ and $\Delta\lambda_{01}$ indicates that the center of the oxide front is elliptically shaped.
In both figures we can clearly distinguish three regimes (I, II, III) which will be discussed later.

We also scan over the micropillar and monitor the reflectance using a 1064 nm Nd:YAG laser.
We use such a wavelength at the red side of the photonic stopband, since here the reflectance spectrum is sensitive to interference of the reflections from the top and bottom DBR mirrors.
As a consequence, changes in the optical length of the spacing layer due to the decrease of the effective refractive index during the oxidation change the reflectance.
This technique therefore allows to probe which regions have a similar oxide thickness and can also be used to monitor in real time the oxide penetration towards the micropillar center \cite{Bakker2013}.

\begin{figure*}[ht!]
\centering
\centerline{\includegraphics[angle=0]{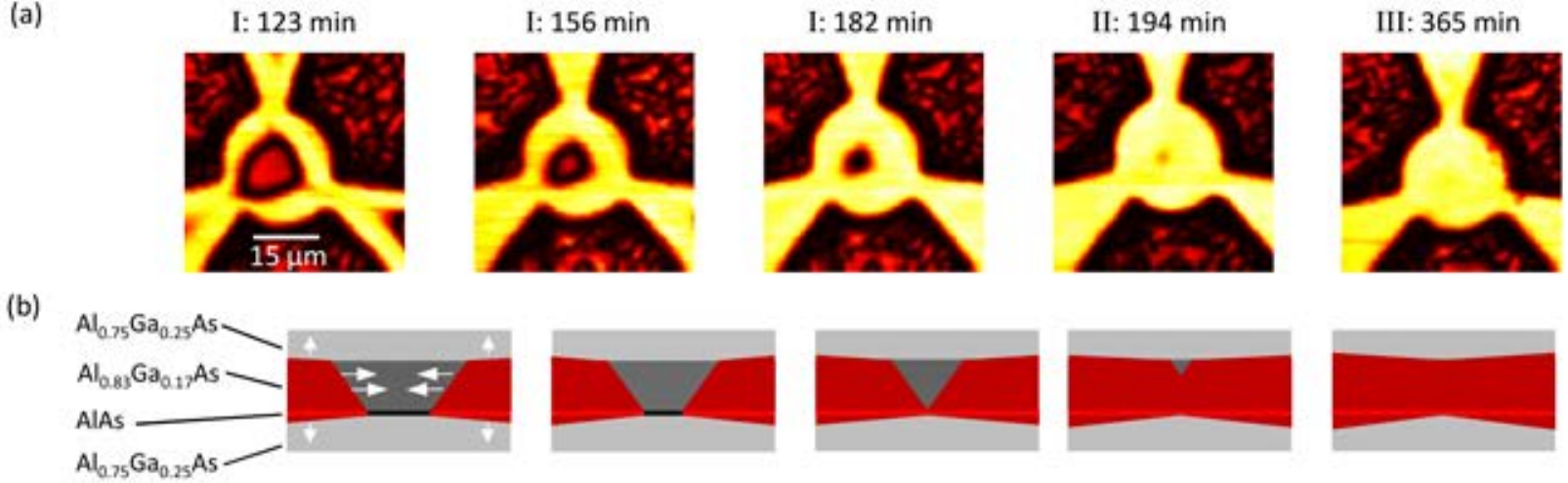}}
\caption{(a) Reflection measurements taken at the red side of the DBR stop band using a focused 1064 nm laser that was scanned across the micropillar center. Light (dark) corresponds to a high (low) reflectance. Regions with a similar reflectance have the same amount of oxide in the aperture region. The title of every scan is the regime number and the total time of all the preceding repetitive oxidations (see main text).
(b) Schematic of the aperture region with the oxide shown in red. We deduce the oxide extent in the different layers from the reflection measurements in (a) and by analyzing the optical modes. The white arrows indicate that the oxidation proceeds much faster horizontally than vertically on this scale, which is largely expanded in vertical direction.}
\label{Fig5}
\end{figure*}

Fig. \ref{Fig5}(a) shows several reflectance images taken after different oxidation steps and the sample was cooled down to room temperature.
The unoxidized parts have a medium reflectance and it can be seen how the front of the oxide aperture, having a lower reflectance, penetrates towards the micropillar center.
Further away from the oxide front the aperture has a high reflectance of constant amplitude, indicating a constant oxide thickness.
When the DBR layers are oxidized this typically also results in a high reflectance, but for this sample and the used oxidation conditions these layers are oxidized less than 2 $\mu$m inwards and are therefore difficult to distinguish.

We support these findings with a SEM image of the FIB cross section from the outward sidewall of the etched trenches to the bulk material, that is presented in Fig. \ref{Fig1}(b).
An oxide aperture with a more or less constant thickness and a $3\mu$m taper is visible.
Also it is seen that the DBR layers are oxidized less than 2 $\mu$m inwards and therefore, in contrast to previous work \cite{Bennet2007}, the oxidation of these layers provides no additional optical confinement.
We therefore schematically display only the oxide extent in the aperture region corresponding to the 1064 nm reflectance images, in Fig. \ref{Fig5}(b).

Through the characterization of the buried oxide layer and the optical modes we are able to identify three regimes (marked with vertical lines at 185 mins and at 203 mins in Fig. \ref{Fig3} and Fig. \ref{Fig4}), that we will discuss now.

I) - The center of the micropillar is initially unoxidized and the oxide taper starts to penetrate inward.
Only after six oxidation steps and a total oxidation time of 150 min has the oxide penetrated far enough that confined Hermite-Gaussian modes can be identified.
The modes start to blueshift due to increasing optical confinement.

II) - The AlAs layer has oxidized all the way to the center of the micropillar and also the oxide taper in the Al$_{0.83}$Ga$_{0.17}$As layer starts to penetrate inward.
During this process a gradual blueshift of up to 30 nm is visible.
Also a steady increase of the mode splitting and thus of the optical confinement continues until a maximum mode splitting of 4 nm is achieved, but then decreases again.
This can indicate that the taper is not perfectly linear but has a larger slope in the center and is flatter at the start and the end of the taper.

III) - The AlAs and Al$_{0.83}$Ga$_{0.17}$As layers in the oxide taper have been fully oxidized, but the oxidation in vertical directions still continues in the surrounding Al$_{0.75}$Ga$_{0.25}$As layers aided by proximity enhancement effects \cite{Blum1997}.
The oxidation goes much slower, since it relies on the smaller oxidation rate of Al$_{0.75}$Ga$_{0.25}$As, but still gives rise to a gradient in the oxide thickness in the horizontal direction and thereby leads to confined modes.
Surprisingly, this gradient is not sufficiently large to show up in the 1064 nm reflection scan.

To gain a quantitative understanding of the blueshift of the cavity modes, which we observe to be 32 nm after 205 min, we use the effective refractive index model \cite{Hadley1995}.
For the refractive index of the oxide we use $n = 1.53$.
Since the 10 nm AlAs layer is located at a node in the field only a 2.5 nm shift should occur when it is oxidized.
The Al$_{0.83}$Ga$_{0.17}$As is thicker and partly located at an antinode, giving a calculated shift of 24.9 nm.
The Al$_{0.75}$Ga$_{0.25}$As below the AlAs layer would give an additional shift of 15.5 nm if fully oxidized.
However it is likely that only partial oxidation of this layer occurred, thereby contributing to the wavelength shift and the confining optical potential still present after 205 min.

Assuming a quadratic confining potential is present, the relationship between the waists of the modes and the mode splitting is given by: $\frac{\Delta\lambda_{10,01}}{\lambda_{00}} = \frac{1}{2}(\frac{\lambda_{00}}{n \pi w_{10,01}})^2$, where $\Delta\lambda_{10} / \Delta\lambda_{01}$ are the mode splittings between the $\Psi_{10}/\Psi_{01}$ modes and the $\Psi_{00}$ mode, $\lambda_{00}$ is the wavelength of the $\Psi_{00}$ mode, $n\approx3.3$ is the average refractive index estimated from the device structure and $w_{10}/w_{01}$ are the waists of the elliptically shaped fundamental mode.
From the mode waists one can calculate the mode area $A = \frac{\pi}{4} w_{10}w_{01}$ and the cavity volume $V = L_{eff}*A$, where $L_{eff}\approx1.4$ $\mu$m is the effective cavity length estimated from the device structure.
At the end of regime I: $\Delta\lambda_{10} = 1.9$ nm, $\Delta\lambda_{01} = 2.3$ nm and $\lambda_{00} = 964$ nm. From this we find $A = 2.1$ $\mu$m$^2$ and $V=2.9$ $\mu$m$^3$.

The quality factor $Q$ of the fundamental optical mode is found by fitting a Lorentzian reflection dip to the reflectance spectra.
At the end of regime I we measure quality factors as high as $Q = 14,000$, but this gradually decreases due to increased scattering by the oxide to $Q = 8,000$ at the end of regime II and decreases further to below $Q = 5,000$ after $t = 350$ mins.

An important figure of merit of microcavities is the Purcell factor $P = \frac{3}{4\pi^2}(\frac{\lambda_{00}}{n})^3\frac{Q}{V}$.
At the end of regime I where the quality factor is at a maximum, the estimated Purcell factor would be $P\approx12$.
This would be an optimal point to aim for.

In conclusion we demonstrated it is possible to repeat the oxidation of a micropillar multiple times, every time adding an extra bit of oxide to the aperture.
This technique can be applied to tune the cavity wavelength more than 30 nm or to obtain mode splittings of up to 4 nm.
In general this technique can be used to obtain desired optical properties with high accuracy.

This work was supported by NSF under Grant No. 0960331 and 0901886 and FOM-NWO Grant No. 08QIP6-2.

\end{document}